\documentclass[aps,prl,showpacs,twocolumn]{revtex4}

\usepackage{amsmath} 
\usepackage{amssymb} 
\usepackage{graphicx}
\usepackage{xspace}

\newcommand{\p}{\partial} 
\newcommand{\bx}{{\bf x}}

\newcommand{\mct}{MC\xspace}
\newcommand{\nequ}{non-equilibrium\xspace}

\def\nbR{\ensuremath{\mathrm{I\!R}}} 
\begin{document}
\draft
\preprint{XXXX}
\title{Universality classes of  the Kardar-Parisi-Zhang equation }
\author{L. Canet,$\,^1$ and  M. A. Moore$\,^2$}
\affiliation{$\,^1$Service de Physique de l'\'Etat Condens\'e,  CEA Saclay, 91191 Gif-sur-Yvette, France\\
 $\,^2$School of Physics and Astronomy, University of Manchester,
Manchester, M13 9PL, UK}

\begin{abstract}

 We re-examine mode-coupling  theory for the Kardar-Parisi-Zhang (KPZ)
 equation in the strong coupling  limit and show that there exists two
 branches of solutions. One branch (or universality class) only exists
 for  dimensionalities $d<d_c=2$  and is  similar to  that found  by a
 variety of  analytic approaches, including  replica symmetry breaking
 and Flory-Imry-Ma  arguments. The second branch exists  up to $d_c=4$
 and gives values  for the dynamical exponent $z$  similar to those of
 numerical studies for $d\ge2$.

\end{abstract}

\maketitle

The celebrated Kardar-Parisi-Zhang  (KPZ) equation \cite{kardar86} was
 initially derived as a model  to describe the kinetic roughening of a
 growing  interface and  has been  the subject  of a  great  number of
 theoretical  studies \cite{halpin95}.  This  is because  the original
 growth problem  has turned  out to be  equivalent to  other important
 physical  phenomena,  such as  the  randomly  stirred fluid  (Burgers
 equation) \cite{forster77}, directed  polymers in random media (DPRM)
 \cite{kardar87}, dissipative transport \cite{beijeren85,janssen86} or
 magnetic  flux  lines   in  superconductors  \cite{hwa92}.   The  KPZ
 equation  has thus  emerged  as one  of  the fundamental  theoretical
 models  for  the  study  of  universality classes  in  \nequ  scaling
 phenomena and phase transitions \cite{halpin95}.

 It   is  a   non-linear   Langevin  equation   which  describes   the
 large-distance, long-time dynamics of the growth process specified by
 a  single-valued  height  function  $h(\bx,t)$ on  a  $d$-dimensional
 substrate $\bx \in \nbR^d$:
\begin{equation}
\p_t h(\bx,t)  = \nu\,\nabla^2 h(\bx,t)  \, + \,\lambda/2\,\big(\nabla
h(\bx,t)\big)^2 \,+\,\eta(\bx,t),
\label{eqkpz}
\end{equation}
where $\eta(\bx,t)$  is a zero  mean uncorrelated noise  with variance
$\langle       \eta(\bx,t)\eta(\bx',t')\rangle       =       2       D
\delta^d(\bx-\bx')\,\delta(t-t')$.    This   equation   reflects   the
competition between  the surface tension  smoothing force $\nu\nabla^2
h$,  the preferential  growth along  the local  normal to  the surface
represented by the non-linear term and the Langevin noise $\eta$ which
tends to roughen the interface and mimics the stochastic nature of the
growth.

The stationary interface is characterized by the two-point correlation
function    $C(|\bx-\bx'|,t-t')    \equiv    \langle    [h(\bx,t)    -
h(\bx',t')]^2\rangle$ and,  in particular, its  large-scale properties
where  $C$  is  expected  to  assume the  scaling  form  $C(d,\tau)  =
d^{2\chi}\,f(\frac{\tau}{d^z})$, where $d=|\bx -\bx'|$ , $\tau=|t-t'|$
and  $\chi$  and  $z$   are  the  roughness  and  dynamical  exponents
respectively.   These  two exponents  are  not  independent since  the
Galilean   symmetry  \cite{forster77}  ---   the  invariance   of  Eq.
(\ref{eqkpz})  under an  infinitesimal  tilting of  the interface  ---
enforces the  scaling relation $z+\chi  = 2$ for  solutions associated
with any fixed point at which $\lambda$ is non-zero.

While some  exact results are available in  $d=1$, yielding $\chi=1/2$
  and $z=3/2$ \cite{kardar87,hwa91,halpin95},
the complete  theoretical understanding of the KPZ  equation in higher
dimensions  is  still  lacking.
 For  $d>2$,  there  exists  a  phase
transition  between two  different  regimes, separated  by a  critical
value      $\lambda_c$     of      the      non-linear     coefficient
\cite{kardar86,forster77}.   In the  weak-coupling regime  ($\lambda <
\lambda_c$), the behavior is governed  by the $\lambda= 0$ fixed point
--- corresponding    to   the   linear    Edwards-Wilkinson   equation
\cite{halpin95} --- with exponents $\chi  = (2-d)/d$ and $z=2$. In the
strong-coupling   (rough)   regime   ($\lambda  >   \lambda_c$),   the
non-linearity becomes  relevant and despite  considerable efforts, the
statistical properties  of the  strong-coupling (rough) regime  for $d
\ge 2$ remain controversial. 

The existence of a finite upper critical dimension $d_c$ at
 which the dynamical exponent 
$z$ of the 
strong-coupling phase becomes 2
   is also  much  debated.   Most  of  the
 analytical approaches  support a finite $d_c$,  but their predictions
 are  varied:  one of the solutions of the mode-coupling  (MC) equations 
 and  other  arguments
 indicate $d_c\simeq 4$ \cite{bouchaud93,halpin89,colaiori01, colaiori01b},
 whereas
 functional   renormalization  group   to   two-loop  order   suggests
 $d_c\simeq  2.5$ \cite{ledoussal03}. A set of 
  related theories such as  replica  symmetry
  breaking  \cite{MP},   variational  studies  \cite{GO}  and
  Flory-Imry-Ma arguments  \cite{MG}  give $d_c=2$.
 On the  other  hand, numerical
 simulations and real space calculations find no evidence at all for a
 finite $d_c$ \cite{tang92,marinari00,castellano98}!
 
 We re-examine  in this  paper \mct theory.   It is  a self-consistent
 approximation \cite{beijeren85,janssen86},  where in the diagrammatic
 expansion for  the correlation  and response functions  only diagrams
 which  do  not  renormalize  the  three-point  vertex  $\lambda$  are
 retained.    The   \mct  approximation   has   been  widely   studied
 \cite{hwa91,frey96,bouchaud93,doherty94,colaiori01}.  Furthermore the
 \mct equations  are exact  for the large  $N$-limit of  a generalised
 $N$-component KPZ  \cite{doherty94}. This  allows in principle  for a
 systematic expansion in $1/N$.
 We shall  show that  within the \mct  approximation there  exists two
  solutions,  one of which had  been previously
  overlooked.  The new solution (universality class)
 only  exists when $d<2$ and seems
  similar to the  solution found in the analytical  studies which give
  $d_c=2$ \cite{MP, GO, MG}.   The other \mct solution exists 
 on the whole range $0<d<4$ and has $d_c=4$.  This previously known
  solution gives in $d=2$ a value for $z$ 
  close  to that from numerical work, e.g.   \cite{marinari00}.

The correlation and response functions are defined in Fourier space by
 $C({ k},\omega)=\langle h({ k},\omega)  h^{*}({ k},\omega)
\rangle$
 and  $G({ k},\omega)\delta^{d}({  k}-{ k'})\delta
  (\omega-\omega')=\Big\langle \displaystyle\frac{\partial   h({   k},\omega)}
 {\partial  \eta({ k'},\omega')} \Big\rangle$,
where $\langle  \cdot \rangle$ indicates  an average over  $\eta$.
In  the  MC   approximation,  the  correlation  and  response
functions are the solutions of two coupled equations,
\begin{eqnarray}
G^{-1}({     k},\omega)&=&    G^{-1}_0({    k},\omega)+\lambda^2
\displaystyle\int       \displaystyle       \frac{d\Omega}{2      \pi}
\displaystyle\int   \displaystyle   \frac{d^dq}{(2  \pi)^d}\, {  q}\cdot
  ({ k}-{  q}) \nonumber\\
&\times &\left({  q}
\cdot  { k}\right)  G({ k}-{  q},\omega  - \Omega)\, C({ q},\Omega) 
\label{mc1}
\\    C({     k},\omega)&=&    C_0({    k},\omega)+\displaystyle
\frac{\lambda^2}{2}  \big| G({  k},\omega)\big|^2 \displaystyle\int
\displaystyle  \frac{d\Omega}{2  \pi} \displaystyle\int  \displaystyle
\frac{d^dq}{(2 \pi)^d}\nonumber\\  &\times &\left({ q}  \cdot ({
k}-{  q})\right)^2 C({  k}-{  q},\omega -  \Omega)\, C({ q},\Omega) 
\label{mc2}
\end{eqnarray}
\noindent
where  $G_0({ k},\omega)=(\nu  k^2 -  i \omega)^{-1}$  is  the bare
response   function,   and   $C_0({   k},\omega)=2   D   |   G({
k},\omega)|^2$  \cite{bouchaud93}.   In  the  scaling  limit,  $G({
k},\omega)$ and $C({ k},\omega)$ take the  forms
 $G({ k},\omega)= k^{-z}g\left( \omega/k^{z}\right)$ 
and $C({ k},\omega)= k^{-(2  \chi+d+z)}n\left( \omega/k^{z}\right)$
 and Eqs. (\ref{mc1}) and  (\ref{mc2}) translate into coupled equations
for the scaling functions $n(x)$ and $g(x)$:
\begin{eqnarray}
g^{-1}(x)&=& -i x +I_{1}(x) \,,
\label{g}
\\ n(x)&=& | g(x)|^{2}I_{2}(x) \,,
\label{n}
\end{eqnarray}
where $x=\omega/k^z$ 
and $I_1(x)$ and $I_2(x)$ are given by \cite{bouchaud93}
\begin{eqnarray}
I_1(x)&=&     P\displaystyle\int_{0}^{\pi}d\theta     \sin^{d-2}\theta
\displaystyle\int_{0}^{\infty}dq\cos\theta    (\cos\theta-q)\nonumber\\
&&\times     q^{2z-3}r^{-z}     \displaystyle\int_{-\infty}^{\infty}dy
\,g\left(\displaystyle \frac{x-q^{z}y}{r^{z}}\right)n(y), \nonumber \\
I_2(x)  &=&\displaystyle \frac{P}{2}\displaystyle\int_{0}^{\pi}d\theta
\sin^{d-2}\theta       \displaystyle\int_{0}^{\infty}dq(\cos\theta-q)^2
\nonumber\\                &&\times               q^{2z-3}r^{-(d+4-z)}
\displaystyle\int_{-\infty}^{\infty}dy           \,n\left(\displaystyle
\frac{x-q^{z}y}{r^{z}}\right)n(y), \nonumber
\end{eqnarray}
\noindent
with         $P\!=\!\lambda^2/(2^d\Gamma(\frac{d-1}{2})\pi^{(d+3)/2})$,
$r^2=1+q^2-2q\cos\theta$.   Notice that  the  ``bare term''  $C_0({
k},\omega)$ has been dropped from  these equations as it is negligible
in the {\em scaling} limit provided $2 \chi+d+z > 2z$, i.e.  $4+d> 3z$.

All  the  (necessarily  approximate)  solutions of  the  MC  equations
involve  an ansatz on  the form  of the  scaling functions  $n(x)$ and
$g(x)$. The  relation $z=z(d)$ is then  obtained requiring consistency
of  Eqs.   (\ref{g})  and  (\ref{n})  on matching  both  sides  at  an
arbitrarily  chosen  value  of   $x$.   Due  to  the  non-locality  of
Eqs. (\ref{g})  and (\ref{n}), the  matching condition depends  on the
form of the functions $n$ and $g$  for all $x$, so the ansatz needs to
be  reliable for  all $x$,  and in  particular satisfy  the  large $x$
asymptotic forms
 $n(x)\sim     x^{-1-\beta/z}$,  $g_R(x)\sim    x^{-1-2/z}$
 and $g_I(x)\rightarrow x^{-1}$
 where $\beta=d+4-2z$ and $g(x)=g_R(x)+ig_I(x)$.
Colaiori and  Moore \cite{colaiori01,colaiori01b} proposed  an ansatz
 which satisfies these large  $x$ constraints. It
  enabled the authors to provide numerical estimates of $z$ when
 $d=2,3$ in reasonable  agreement with exponents  obtained from
 simulations  \cite{marinari00,tang92}.  It  also  yielded an  integer
 finite upper critical dimension $d_c=4$ \cite{colaiori01}.  Moreover,
 it led in $d=1$ to the discovery of a stretched exponential behavior
 for the two-point  correlation function \cite{colaiori01b}   similar to
  that found in the exact solution \cite{PS04}.

 Using a similar  ansatz, we next show  that  Eqs.  (\ref{g})  and
 (\ref{n}) have an additional solution with $d_c=2$, before discussing
 its significance. The advocated ansatz is most
 conveniently expressed in Fourier space,
\begin{eqnarray}
\widehat{g}(p) &= &\theta(p) \,\exp(- | p|^{2/z}) \,,
\label{gansatz}\\
  \widehat{n}(p)&=&A\exp(-
\displaystyle\big|B p\big|^{\beta/z}) \,,
\label{nansatz}
\end{eqnarray} 
where  $\widehat{n}$ and  $\widehat{g}$ are  the Fourier  transform of
 $n(x)$   and   $g(x)$   respectively,   $A$ is a (nonzero) parameter
 and $B\equiv B(d,z)$ an arbitrary function
 \cite{colaiori01}.   Requiring  the matching  of  Eqs. (\ref{g})  and
 (\ref{n}) at large  $x$ then yields:
\begin{eqnarray}
{C_d\,A}\,{B}^{-1}& =&  {d\,(2-z)}/{(z^2\,I(B,d,z))} \label{eq1}\\
{C_d\,A}\,{B^{-2}} & =& (\beta/z)^2\,(\beta-z)\,{2^{\frac{2z-\beta}{\beta}}}/{\Gamma\big( \frac{2z-\beta}{\beta}\big)}\label{eq2}
\end{eqnarray}
where $C_d =  \lambda^2\,\Gamma((d-1)/2)/(4\,\pi^{\frac{d+1}{2}}\,\Gamma(d-1))$ and
$  I(B,d,z)\!=\!\displaystyle\int_{0}^{\infty}\!ds  \,  B\,  (1-2s^2)s^{2z-3}
\exp\big(-B^{\beta/z}s^{\beta}-s^2\big)$.
Combining  (\ref{eq1}) and (\ref{eq2}) then gives an implicit equation for the curve $z(d)$ depending
on $B$ only:
\begin{equation}
\beta^2\,(\beta-z)\,\frac{2^{(2z-\beta)/\beta}}{\Gamma\big(\frac{2z-\beta}{\beta}\big) }\,I(B,d,z) = \frac{d\,(2-z)}{B}.
\label{eqmct}
\end{equation}
 Eq. (\ref{eqmct})  can be solved numerically once $B(d,z)$ has been fixed.
 Prior to  discussing  the choice of $B$, 
  we stress the ($B$-independent)  
 intrinsic properties of the  solutions $S_{d,z}$.
First, as $z\to 2$,  $I(B,d,2)$ is finite for any  $B>0$.
One can then infer from  Eq. (\ref{eqmct}) that $S_{2,2}$ and $S_{4,2}$
are always solutions independent of $B$ (provided $B^{-1}$ is finite).
This  singles out the two  dimensions $d_c = 2$ and $d_c=4$ as critical ones
  since $z\to 2$ for $d\to d_c$.
 Second, as $d\to 0$,  $I(B,0,z)$ is non-zero for any
   $B>0$, and  Eq. (\ref{eqmct}) yields the two
 solutions $S_{0,4/3}$ and $S_{0,1}$  (provided again
 $B^{-1}$ is finite).
Lastly, as the MC equations satisfy  the fluctuation-dissipation theorem (FDT)
 that exists in $d=1$ \cite{frey96} yielding the exact result $z=3/2$,
  we expect the ansatz to preserve this property,
  which is fulfilled for
 $B(1,3/2)=1$ as can be checked on Eqs. (\ref{gansatz}) and (\ref{nansatz}).
 It follows that  $S_{1,3/2}$ is  solution of  (\ref{eqmct}).
 To summarize, Eq. (\ref{eqmct}) has  five  intrinsic solutions
  $S_{d,z}$ which do not depend on the choice of $B$
   (provided  $B^{-1}<\infty$  and $B(1,3/2)~=~1$).

 Furthermore,   Eq. (\ref{eq1})  provides  additional  constraints on  $B(d,z)$
  in the two limits  $z\to 2$ and $d\to 0$.
As $I(B,d,2)$ is finite as $z\to 2$,
  the right-hand side of (\ref{eq1}) vanishes identically,
 which forces $B^{-1}\sim(2-z)$ on the left-hand side.
 One can then derive  the  expansions of two solutions, labelled $z_{\rm S}$ and $z_{\rm F}$,
  about  their critical  dimensions:
\begin{eqnarray}
z_{\rm F} &=& 2 - (4-d)/4 + {\cal O}((4-d)^2) \label{eps4} \\
z_{\rm S} &=& 2 - (2-d)/3 + {\cal O}((2-d)^2). \label{eps2}
\end{eqnarray}
 The arguments of Ref.  \cite{colaiori01} can be used to show
 that these expansions are  very robust and more general
 than  the particular ansatz of  Eqs.  (\ref{gansatz}) and
(\ref{nansatz}).
Similarly, 
 as $I(B,0,z)$ is finite for $d\to 0$ and $z>1$, Eq. (\ref{eq1})
  imposes  $B^{-1}\sim d$.
 The  expansions  of the two solutions about $d=0$ can also  be worked out: 
$z_{\rm F} = 4/3+d/3 + {\cal O}(d^2)$ and 
$z_{\rm S} = 1 + d/4 + {\cal O}(d^2)$.

\begin{figure}[ht]
\includegraphics[height=80mm,angle=-90]{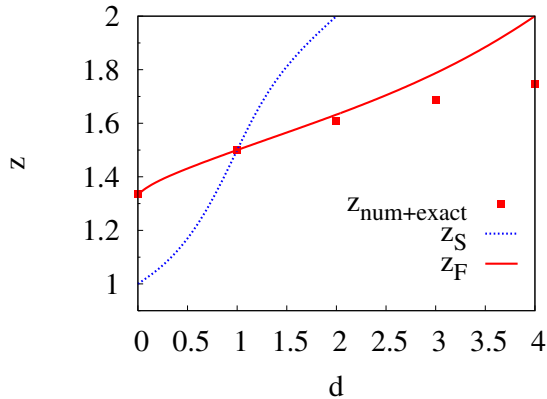}
\caption{The two  solutions $z_{\rm S}(d)$  and $z_{\rm F}(d)$  of the
 \mct  equations   found  in   this  work,  compared   with  numerical
 \cite{marinari00} results for $d=2,3,4$ and the exact results in $d=0,1$,
 all represented by squares.}
\label{z}
\end{figure}

 We emphasize that all these  
 features 
 do not depend on the choice of $B$.
  Any ansatz would lead
  to the same qualitative picture as Fig. \ref{z} provided  $B\sim (d(2-z))^{-1}$. 
 The specific one  used here further exploits 
 the {\it uniqueness} of the solution in $d=1$ which imposes additional
constraints on $B$ \footnote{ Uniqueness
 of 
 the solution in $d=1$  requires $S_{1,3/2}$ to be an extremum of
 the implicit curve $z(d)$, which in turn yields:
$ \alpha_d\equiv\frac{\partial B}{\partial d}(1,3/2)= \frac{1}{56} (-60+21 \gamma +7 \log (512))$
 and $\alpha_z\equiv\frac{\partial B}{\partial z}(1,3/2)= -\frac{3}{28} (-36+7 \gamma +21 \log (2))$.
 Our final ansatz for $B(d,z)$ then reads
$B(d,z) = (1 + (\alpha_d+1)(d-1)+(\alpha_z-2)(z-3/2))/(2\,d\,(2-z)).$
}.
 With this choice, we have solved Eq. (\ref{eqmct})  numerically in 
 $0<d<4$
  and found that there  exists two branches
 of solutions  $z_{\rm S}(d)$  and $z_{\rm F}(d)$
 lying  in the  interval  $d\in  (0,2)$ but  only  one solution,  the
 continuation of $z_{\rm  F}$, in the interval $d\in  (2,4)$.
%
%

The  branch $z_{\rm F}$, referred  to as the F  solution as it
 exists over  the \textit{full}  dimensionality range $d\in  (0,4)$, is 
  similar to the 
 solution  analyzed in  \cite{colaiori01} \footnote{Note that in  \cite{colaiori01}
 $B=(2(2-z))^{-1}$ is used instead of [26]. Most features
  of  the F and S solutions 
 with both choices are very similar; the $z_F(d)$ curves
 lie within less than $1\%$ for $d\ge 2$. However, only the latter
   leads to continuous solutions
 on either side of $d=1$.}. 
   It lies  fairly close  to
 typical numerical  results \cite{marinari00} with  $z=1.632$ for $d= 2$ and $z=1.787$ for
 $d=3$, reproduces the exact results 
 in  $d=0$ and $d=1$ (see Fig.  \ref{z})  and reaches $z=2$ at $d_c=4$,  which defines the
 upper critical dimension  of this solution.
         The  other branch,  denoted  S as  it  is associated  with the 
 \textit{shorter}
 dimensionality interval  $d\in (0,2)$, embodies a new universality class
 with  $z_{\rm S} \neq z_{\rm F}$  for $d\neq 1$. 
  For  this solution,  $z=2$ at  $d_c=2$ and it does  not exist
 for $d\ge 2$, so that there
 remains a single strong-coupling solution in and above two dimensions
 -- the F solution.
 We emphasise that both the F and S solutions are non-perturbative since
 no (stable) perturbative fixed point can be found
 for $d<2$ within the standard perturbative expansion 
 in  the coupling constant  $g={2\lambda^2D}/{\nu^3}$, even to all orders
  \cite{Wiese}.  \mct  theory involves  resumming an  infinite
 set of diagrams and  so is capable of going beyond standard RG techniques.

 We now comment on our findings. 
 According to MC theory,  for $d < 2$ there are two possible
 strong-coupling universality  classes,  F and S. 
It is clear that for  a given set  of initial conditions and
 bare correlators, the time evolution would unambiguously steer the system
 towards one or other  of the two solutions.  
Equivalently, in the mapping to  the DPRM,
the solution with  lowest free energy would  be selected as the physical one.
 The  apparent  freedom  to  choose a  solution  arises  
 because we have considered the scaling regime in which the bare 
 correlators have been dropped as  irrelevant, so we have lost track
 of their role. However, a clue as to the form of bare correlator 
 needed for the S solution to emerge can be obtained by inspecting the zero-dimensional case.
 In $d=0$, the  DPRM analogue  is a  directed walk  along a  chain of
length  $L$ on which  the  site  energies  are random.   A  dynamical
exponent $z=1$  would correspond to free energy  fluctuations of order
$L^{2/z-1}=L$  which  is  larger  than the  $\sqrt{L}$  (for
$z=4/3$) expected  for independent  random site energies.  
with $z=1$  could only exist with long-range  correlations between the
site energies  and is therefore not the  natural solution  for the short-range problem in  $d=0$. We suspect
that the $z_{\rm S}$ branch in the whole range $0<d<2$ would only be realized
 if  there were some  long-range correlations in the noise for  the KPZ
problem or between the site energies for the DPRM analogue.

To investigate this possibility further,
 suppose  instead   of  starting  out  with   noise  with  short-range
  correlations, one introduces  some long-ranged correlated noise, of
  the form  $\langle \eta({ x},t)\eta({  x'},t')\rangle = 2 R({
  x}-{  x'})\delta(t-t')$ with some power law distribution $R({  x}) \propto  x^{2\rho-d}$, or
  equivalently   in  Fourier   space  $R({   k}) =   D  (1+w
  k^{-2\rho})$. The MC equations would formally be the same with 
  $D$ replaced by  the  `generalized' noise correlator
 $R({ k})$ \cite{frey96}.
  These long-range bare noise correlations will not affect
  the  scaling  behavior  of  the  system  as  long  as  they  decay
  sufficiently  quickly,  so  that  they  become irrelevant after  some
  transient regime. The condition for this to happen can be worked out
  by substituting  the scaling forms for $G$  and $C$ into
  Eq. (\ref{mc2}). The initial correlations decay as
$C_0({ k},\omega)=2  R({ k})|G({
k},\omega)|^2 \sim k^{-2(z+\rho)}$
 whereas  the long-time correlation function is expected to behave
 as $C({ k},\omega)\sim
 k^{-(2\chi +d+z)}\equiv k^{-(d+4-z)}$ using the identity $z+\chi=2$.
 Comparing these two expressions, one infers that the long-range correlations
 will  destabilize a short-range  solution with dynamical exponent $z$ when
\begin{equation}
\rho > \frac{d+4-3z}{2},
\label{cond}
\end{equation}
which is identical to the  criterion obtained in \cite{frey99} from RG
 arguments.  The condition of Eq. (\ref{cond})   determines the
 phase  boundary  $\rho(d,z)$  between  the short-range  and  long-range
 stability  domains for  each  of  the two  \mct  solutions $z_{\rm F}$  and
 $z_{\rm S}$. In the long-range noise dominated region
$z_{\rm LR}=\frac{d+4-2\rho}{3}$ \cite{frey99}. The solutions F and 
S are stable against  noise of range $\rho$ if their associated $z$ satisfies $z<z_{\rm LR}$. 
 For all $0<d<2$, we believe that 
the S solution requires for its {\em existence}
the addition of  long-range noise of infinitesimal amplitude with
 $\rho_S= \frac{d+4-3z_{\rm S}}{2}$, as in the case $d=0$ (where $z=1$ would correspond
 to long-range noise with $\rho=0.5^{-}$). For $0<d<1$, the S solution  lies at the stability limit
of the long-range solution and in the interval
$1<d<2$, it  will be the solution to the KPZ problem  in the presence of
infinitesimal amounts of such correlated noise if (say) the associated free energy
in the DPRM analogue is lower than that of the F solution (with the same correlated noise).
 This S solution could hence possibly be brought out by tuning the  parameters.
 We believe that
in the absence of  some long-range correlation, the F solution will be the appropriate solution
 or universality class
 for  $0<d<4$.

Our S solution might  appear to be of academic interest as  it only exists
when $d<2$ and coincides with the F one in $d=1$. However,
 it turns out to be strikingly  similar to the solution found in several theoretical studies.
We focus  on three of them: the replica  symmetry breaking (RSB)
  calculation   of  M\'{e}zard   and  Parisi   \cite{MP},   a  Gaussian
  variational calculation  \cite{GO} and Flory-Imry-Ma  (FIM) arguments
 \cite{MG}.  All  three   approaches  result  in  a  strong-coupling
  solution  with   $d_c  =2$.  The  RSB
   treatment yields a dynamical exponent equivalent to the FIM
 result \cite{MP}: 
 $z_{\rm FIM}=(4+d)/3   \simeq  2-(2-d)/3   +  {\cal
O}((2-d)^2)$,
 which is  identical to (\ref{eps2}) and  suggests that the \mct  solution S and
 the  RSB --  FIM   solution are indeed
 different approximations to the same underlying  universality class. 
It is not apparent how the long-range correlations needed for the
S solution to exist enter  these analytical approaches. However, it has been observed long ago
  that by the addition of a type of long-range force,  the Flory formula in the
conventional self-avoiding polymer problem  becomes exact \cite{MB}.

To summarize, we have found a new \mct solution, S, which has an upper
critical dimension $d_c=2$, in addition to the usual one, F, which has
$d_c=4$. This  new solution seems   similar to  that
 found  by  analytical  treatments  such  as  RSB
   and  FIM  arguments, but we believe will only exist in the
presence of appropriate long-range correlated noise.
  We  acknowledge  that  the
numerical  value for  $z$  shown in  Fig.  \ref{z} when  $d=4$ is  not
apparently consistent with the upper critical dimension $d_c=4$ of the
F  solution, but  we  attribute  that to  the  inability of  numerical
studies in high dimensions to access the scaling regime.

\begin{acknowledgments}
The authors wish  to thank A. Bray, B. Derrida,  T. Garel, C. Monthus,
H.  Orland,  U.C.  T\"auber  and  K.  Wiese  for  useful  discussions.
Financial  support   by  the  European   Community's  Human  Potential
Programme   under    contract   HPRN-CT-2002-00307,   DYGLAGEMEM,   is
acknowledged. One of  us (MAM) would like to thank  CEA Saclay for its
hospitality.
\end{acknowledgments}


\begin{thebibliography}{10}

\bibitem{kardar86}   M.   Kardar,   G.   Parisi,  and   Y.-C.   Zhang,
Phys. Rev. Lett. {\bf 56}, 889 (1986).

\bibitem{halpin95} T. Halpin-Healy and Y. Zhang, Phys. Rep. {\bf 245},
218 (1995), J. Krug, Adv. Phys. {\bf 46}, 139 (1997).

\bibitem{forster77}  D.  Forster,  D.~R.  Nelson, and  M.~J.  Stephen,
Phys. Rev. A {\bf 16}, 732 (1977).

\bibitem{kardar87} M. Kardar, Nucl. Phys. B {\bf 290}, 582 (1987).

\bibitem{beijeren85}  H.  van  Beijeren,  R.  Kutner,  and  H.  Spohn,
Phys. Rev. Lett. {\bf 54}, 2026 (1985).

\bibitem{janssen86} H.~K. Janssen and  B. Schmittmann, Z. Phys. B {\bf
63}, 517 (1986).

\bibitem{hwa92} T. Hwa, Phys. Rev. Lett. {\bf 69}, 1552 (1992).

\bibitem{hwa91}  T. Hwa  and E.  Frey, Phys.  Rev. A  {\bf  44}, R7873
(1991).


\bibitem{bouchaud93} M.~A.  Moore {\it et~al.}, Phys.  Rev. Lett. {\bf
74}, 4257  (1995), J.-P. Bouchaud and  M.~E. Cates, Phys.  Rev. E {\bf
47}, R1455 (1993)

\bibitem{halpin89}  T. Halpin-Healy,  Phys. Rev.  Lett. {\bf  62}, 442
(1989), H.~C. Fogedby, Phys. Rev. Lett. {\bf 94}, 195702 (2005).

\bibitem{colaiori01}     F.      Colaiori     and     M.~A.     Moore,
Phys. Rev. Lett. {\bf 86}, 3946 (2001).

\bibitem{colaiori01b} F.  Colaiori and M.~A. Moore, Phys.  Rev. E {\bf
63}, 057103 (2001), Phys. Rev. E {\bf 65}, 017105 (2001).

\bibitem{ledoussal03} P.  {Le~Doussal} and  K.~J. Wiese, Phys.  Rev. B
{\bf 68}, 174202 (2003), Phys. Rev. E {\bf 72}, 035101(R) (2005).

\bibitem{MP}  M. M\'{e}zard  and G.  Parisi, J.  Phys. I,{\bf  1}, 809
(1991).

\bibitem{GO}  T. Garel  and  H. Orland,  Phys.  Rev. B  {\bf 55},  226
(1997).

\bibitem{MG} C.  Monthus and T. Garel,  Phys. Rev. E  {\bf 69}, 061112
(2004).


\bibitem{tang92}   L.-H.  Tang,   B.~M.  Forrest,   and   D.~E.  Wolf,
Phys.  Rev.  A  {\bf  45},  7162 (1992),  T.  Ala-Nissila,  T.  Hjelt,
J.~M. Kosterlitz, and O. Ven\"al\"ainen, J. Stat.  Phys. {\bf 72}, 207
(1993),  E.   Marinari,  A.  Pagnani,   G.  Parisi,  and   Z.  R\'acz,
Phys. Rev. E {\bf 65}, 026136/1 (2002).

\bibitem{marinari00}   E.  Marinari,  A.   Pagnani,  and   G.  Parisi,
J. Phys. A {\bf 33}, 8181 (2000).

\bibitem{castellano98} C.  Castellano, M. Marsili,  and L. Pietronero,
Phys. Rev.  Lett. {\bf  80}, 3527 (1998),  C. Castellano,  M. Marsili,
M.~A.  Mu{\~n}oz, and  L.  Pietronero,  Phys. Rev.  E  {\bf 59},  6460
(1999).

\bibitem{doherty94}  J.~P.  Doherty,  M.~A.  Moore,  J.  M.  Kim,  and
A.~J. Bray, Phys. Rev. Lett. {\bf 72}, 2041 (1994).

\bibitem{frey96} E.  Frey, U.~C.  T\"auber, and T.  Hwa, Phys.  Rev. E
{\bf 53}, 4424 (1996).

\bibitem{PS04}  M. Pr\"{a}hofer and H. Spohn, J. Stat. Phys. {\bf 115}, 255 (2004).

\bibitem{Wiese}
K. J. Wiese, J. Stat. Phys. {\bf 93}, 143 (1998).

\bibitem{frey99}
 {E. Frey,  U. C. T\"auber, and H. K. Janssen},  {Europhys. Lett. 47}, 14,
   1999.

\bibitem{MB}
M.~A. Moore and A.~J. Bray, J. Phys. A: Math. Gen. {\bf 11}, 1353 (1978), Y. Chen
and R.~A. Guyer J. Phys. A: Math. Gen. {\bf 21}, 4173 (1988).


 

\end{thebibliography}

\end{document}